\documentclass[12pt]{article}
\usepackage{rotating}
\catcode`\@=11
  
\def\@normalsize{\@setsize\normalsize{15pt}\xiipt\@xiipt
\abovedisplayskip 14pt plus3pt minus3pt%
\belowdisplayskip \abovedisplayskip
\abovedisplayshortskip  \z@ plus3pt%
\belowdisplayshortskip  7pt plus3.5pt minus0pt}
\def\small{\@setsize\small{13.6pt}\xipt\@xipt
\abovedisplayskip 13pt plus3pt minus3pt%
\belowdisplayskip \abovedisplayskip
\abovedisplayshortskip  \z@ plus3pt%
\belowdisplayshortskip  7pt plus3.5pt minus0pt
\def\@listi{\parsep 4.5pt plus 2pt minus 1pt
            \itemsep \parsep
            \topsep 9pt plus 3pt minus 3pt}}
 
\def\underline#1{\relax\ifmmode\@@underline#1\else
        $\@@underline{\hbox{#1}}$\relax\fi}
\@twosidetrue
\relax

\catcode`@=12

\catcode`\@=11

\def\section{\@startsection{section}{1}{\z@}{3.5ex plus 1ex minus
   .2ex}{2.3ex plus .2ex}{\large\bf}}

\def\ps@headings{\def\@oddfoot{}\def\@evenfoot{}
\def\@oddhead{\hbox{}\hfill
        \makebox[.5\textwidth]{\raggedright\ignorespaces --\thepage{}--
        \hfill }}
\def\@evenhead{\@oddhead}
\def\subsectionmark##1{\markboth{##1}{}}
}
 
\ps@headings
 
\catcode`\@=12
 
\relax

\def\figcap{\section*{Figure Captions\markboth
        {FIGURECAPTIONS}{FIGURECAPTIONS}}\list
        {Fig. \arabic{enumi}:\hfill}{\settowidth\labelwidth{Fig. 999:}
        \leftmargin\labelwidth
        \advance\leftmargin\labelsep\usecounter{enumi}}}
 \relax
\def\tablecap{\section*{Table Captions\markboth
        {TABLECAPTIONS}{TABLECAPTIONS}}\list
        {Table \arabic{enumi}:\hfill}{\settowidth\labelwidth{Table 999:}
        \leftmargin\labelwidth
        \advance\leftmargin\labelsep\usecounter{enumi}}}
 \relax
\def\reflist{\section*{References\markboth
        {REFLIST}{REFLIST}}\list
        {[\arabic{enumi}]\hfill}{\settowidth\labelwidth{[999]}
        \leftmargin\labelwidth
        \advance\leftmargin\labelsep\usecounter{enumi}}}
 \relax
 
\catcode`\@=11
 
\def\marginnote#1{}
\newcount\hour
\newcount\minute
\newtoks\amorpm
\hour=\time\divide\hour by60
\minute=\time{\multiply\hour by60 \global\advance\minute by-
\hour}
\edef\standardtime{{\ifnum\hour<12 \global\amorpm={am}%
    \else\global\amorpm={pm}\advance\hour by-12 \fi
    \ifnum\hour=0 \hour=12 \fi
    \number\hour:\ifnum\minute<100\fi\number\minute\the\amorpm}}
\edef\militarytime{\number\hour:\ifnum\minute<100\fi\number\minute}
\def\draftlabel#1{{\@bsphack\if@filesw {\let\thepage\relax
  \xdef\@gtempa{\write\@auxout{\string
    \newlabel{#1}{{\@currentlabel}{\thepage}}}}}\@gtempa
    \if@nobreak \ifvmode\nobreak\fi\fi\fi\@esphack}
     \gdef\@eqnlabel{#1}}
\def\@eqnlabel{}
\def\@vacuum{}
\def\draftmarginnote#1{\marginpar{\raggedright\scriptsize\tt#1}}
\def\draft{\oddsidemargin -.5truein
        \def\@oddfoot{\sl preliminary draft \hfil
        \rm\thepage\hfil\sl\today\quad\militarytime}
        \let\@evenfoot\@oddfoot \overfullrule 3pt
        \let\label=\draftlabel
        \let\marginnote=\draftmarginnote
 
\def\@eqnnum{(\theequation)\rlap{\kern\marginparsep\tt\@eqnlabel}%
\global\let\@eqnlabel\@vacuum}  }
\def\preprint{\twocolumn\sloppy\flushbottom\parindent 1em
        \leftmargini 2em\leftmarginv .5em\leftmarginvi .5em
        \oddsidemargin -.5in    \evensidemargin -.5in
        \columnsep 15mm \footheight 0pt
        \textwidth 250mmin      \topmargin  -.4in
        \headheight 12pt \topskip .4in
        \textheight 175mm
        \footskip 0pt
 
\def\@oddhead{\thepage\hfil\addtocounter{page}{1}\thepage}
        \let\@evenhead\@oddhead \def\@oddfoot{} \def\@evenfoot{}
}
\def\titlepage{\@restonecolfalse\if@twocolumn\@restonecoltrue\onecolumn
     \else \newpage \fi \thispagestyle{empty}\c@page\z@
        \def\thefootnote{\fnsymbol{footnote}} }
\def\endtitlepage{\if@restonecol\twocolumn \else  \fi
        \def\thefootnote{\arabic{footnote}}
        \setcounter{footnote}{0}}  %\c@footnote\z@ }
\catcode`@=12
\relax
\def\ps@headings{\def\@oddfoot{}\def\@evenfoot{}
\def\@oddhead{\hbox{}\hfill
        \makebox[.5\textwidth]{\raggedright\ignorespaces --\thepage{}--
        \hfill }}
\def\@evenhead{\@oddhead}
\def\subsectionmark##1{\markboth{##1}{}}
}
 
\ps@headings
 
\relax
 
\newcommand{\ve}[1]{{\varepsilon}_{#1}}
\newcommand{\cve}[1]{{\varepsilon}_{#1}^{*}}
\newcommand{\Ph}[2]{{\Phi}_{#1#2}}
\newcommand{\ph}[2]{{\phi}_{#1#2}}
\newcommand{\Phb}[2]{{\bar\Phi}_{#1#2}}
\newcommand{\phb}[2]{{\bar\phi}_{#1#2}}
\newcommand{\cPh}[2]{{\bar\Phi}_{#1#2}^{*}}
\newcommand{\cph}[2]{{\bar\phi}_{#1#2}^{*}}
\newcommand{\cPhb}[2]{{\bar\Phi}_{#1#2}^{*}}

\renewcommand{\l}[1]{{\lambda}_{#1}}
\newcommand{\lb}[1]{{\bar\lambda}_{#1}}
\newcommand{\D}[1]{{\Delta}_{#1}}

\newcommand{\si}[1]{\left[{#1}\right]}
\def\firstpage#1#2#3#4#5#6{
\begin{document}
%\draft
%\input epsf.tex
\newcommand{\newc}{\newcommand}
\newc{\ra}{\rightarrow}
\newc{\lra}{\leftrightarrow}
\newc{\beq}{\begin{equation}}
\newc{\be}{\begin{equation}}
\newc{\eeq}{\end{equation}}
\newc{\ee}{\end{equation}}
\newc{\bea}{\begin{eqnarray}}
\newc{\eea}{\end{eqnarray}}
\def\eps{\epsilon}
\def\la{\lambda}
\def\f{\frac}
\def\nm{\nu_{\mu}}
\def\nt{\nu_{\tau}}
\def\sq{\sqrt{2}}
\def\ri{\rightarrow}
\newc{\sm}{Standard Model}
\newc{\smd}{Standard Model}
\newc{\barr}{\begin{eqnarray}}
 \newc{\earr}{\end{eqnarray}}
%%%%%%%%%%%
\begin{titlepage}
\nopagebreak
\title{\begin{flushright}
        \vspace*{-0.8in}
{ \normalsize  hep-ph/9907476 \\
CERN-TH/99-226 \\
  IOA-16/1999 \\
July 1999 \\
}
\end{flushright}
\vfill
{#3}}
\author{\large #4 \\[1.0cm] #5}
\maketitle
\vskip -7mm
\nopagebreak
\begin{abstract}
{\noindent #6}
\end{abstract}
\vfill
\begin{flushleft}
\rule{16.1cm}{0.2mm}\\[-3mm]

%%%%%%%%%%%
%November 1997
\end{flushleft}
\thispagestyle{empty}
\end{titlepage}}
 
\def\simlt{\stackrel{<}{{}_\sim}}
\def\simgt{\stackrel{>}{{}_\sim}}
\date{}
\firstpage{3118}{IC/95/34}
{\large\bf Higgs Mass Textures in Flipped $SU(5)$}
{John Ellis$^{\,a}$,  G.K. Leontaris$^{\,a,b}$
and J. Rizos$^{\,b}$}%\\[-3mm]
{\normalsize\sl
$^a$Theory Division, CERN, CH 1211 Geneva 23, Switzerland\\[2.5mm]
\normalsize\sl
$^b$Theoretical Physics Division, Ioannina University,
GR-45110 Ioannina, Greece\\[2.5mm]
 }
{We analyze the Higgs doublet-triplet mass splitting problem in the
version of flipped $SU(5)$ derived from string theory. Analyzing 
non-renormalizable terms up to tenth order in the superpotential,
we identify a pattern of field vev's that keeps one pair of electroweak
Higgs doublets light, while all other Higgs doublets and all Higgs
triplets are kept heavy, with the aid of
the economical missing-doublet mechanism found in the
field-theoretical version of flipped $SU(5)$. The solution predicts
that second--generation charge -1/3 quarks and charged leptons are
much lighter than those in the third generation.}

\newpage
\section{Introduction}

One of the most challenging problems in supersymmetric GUT
model-building is that of the doublet-triplet mass splitting \cite{DG}.
The
problem is less severe than in non-supersymmetric GUTs, because
supersymmetry can be used to stabilize light masses for electroweak
Higgs doublets, but how did they get to be light in the first place?
And can this be arranged in a natural way that also keeps their GUT
triplet partners sufficiently heavy, safeguarding the stability of the
proton?

One of the most promising approaches to this problem is the
missing-doublet idea~\cite{TDA,TDB}. One postulates a GUT with additional
colour-triplet fields, that couple to the Higgs triplets and give them
GUT-scale masses, but no accompanying electroweak doublet fields, so
that the Standard Model Higgs doublets may remain light. The most
economical realization~\cite{FSU5B} of this idea is in the
field-theoretical 
version of the flipped SU(5)$\times$U(1) GUT, where the unwanted Higgs
triplets pair with triplet fields in the 10-dimensional representations
that develop GUT-scale vacuum expectation values
(vev's)~\cite{FSU5B,FSU5A}.

This was one of the vaunted advantages of flipped SU(5), along with
its lack of adjoint and higher GUT representations, that motivated its
derivation from string theory \cite{aehn}. 
String-derived flipped SU(5) models have
provided one of the more successful avenues in string phenomenology,
but there has never been a definitive answer whether they resolve the
doublet-triplet mass-splitting problem. This is complicated by the fact
that flipped $SU(5)$ models derived from string contain several
$\bf{5}$ and $\bf\overline{{5}}$ Higgs representations with
candidate electroweak Higgs fields, as well as considerable
ambiguity in the assignments of $\bf{10}$ and
$\bf\overline{{10}}$ representations. On the other hand, the
generalized Yukawa couplings of chiral supermultiplets can be
calculated, including (in principle) non-renormalizable
couplings~\cite{NR} up to
any desired order (depending on one's computational stamina).

At the level of renormalizable Yukawa couplings, it is known that there
are pairs of massless electroweak doublets, and some couplings between
Higgs triplets and prospective partners in $\bf{10}$ and
$\bf\overline{{10}}$ representations. Some relevant
non-renormalizable couplings have also been identified~\cite{NR}, but a
systematic survey is lacking. One expects the contributions of such
non-renormalizable interactions to be suppressed by powers of $V/M$,
where $V$ is the generic vev of some singlet or
$\bf{10}$/$\bf\overline{{10}}$ field,
and $M$ is related to the Planck scale. One might expect $M
\sim {\cal O} (M_P/\sqrt{8\pi}) \sim 10^{18}$ GeV in a
weakly-coupled string model, and perhaps $M \sim 10^{16}$ GeV in a
strongly-coupled model. One might expect $V/M = {\cal O}(1/10)$, and
non-renormalizable interactions with such a suppression factor 
have been shown to provide
encouraging textures for quark, charged-lepton 
and neutrino mass matrices. In this paper,
we study whether the string-derived flipped $SU(5)$ model contains
non-renormalizable interactions which, with a suitable pattern of large
vev's, provide an acceptable texture for Higgs masses, with light
doublets and heavy triplets.

We first identify textures of Higgs mass matrices that provide one pair
of light Higgs doublets while keeping all Higgs triplets heavy. Then we
examine all the relevant renormalizable and non-renormalizable Yukawa
interactions up to tenth order in the fields. Starting from
constraints on large vev's $V$
that are imposed by earlier phenomenological studies, we explore
whether there is a suitable refinement of the pattern of vev's that
may solve the doublet-triplet mass-splitting problem in flipped SU(5).
We exhibit a pattern of vev's, i.e., a possible choice of string
vacuum, in which there is one pair of light Higgs doublets that
remain massless through tenth order in the superpotential, another
pair of Higgs doublets and a pair of Higgs triplets that acquire
masses ${\cal O}(10^{10}$~to~$10^{12})$~GeV, and the remaining
Higgs doublets and triplets have masses close to the GUT scale.
 
\section{The Problem of Light Higgs Doublets in Flipped $SU(5)$}

In the minimal field-theoretical version of flipped $SU(5)$, there is a
single pair of electroweak Higgs doublets , which
are separated from their heavy Higgs triplet partners $D_3, {\bar D}_3$
 by an economical realization of the missing-doublet mechanism, in
which $D_3, {\bar D}_3$ couple to triplet members of $\bf{10}$
and $\bf10$ representations $H, {\bar H}$ and
acquire large masses from GUT-scale vev's of electroweak-singlet
components of $H, {\bar H}$ \cite{FSU5B}. The light electroweak Higgs doublets
 couple via a singlet field $\phi$ that develops a
vev at the supersymmetry-breaking scale, providing an acceptable value of
the Higgs mixing parameter $\mu$. 

This simple and elegant
picture becomes more complicated in the version of flipped $SU(5)$
derived from string theory in the free-fermion formulation.
This model is obtained by introducing a
set of five vectors of world-sheet boundary conditions $(1, S,
b_{1,2,3})$,
which define an $SO(10)\times SO(6)\times E_8$ gauge group with N=1
supersymmetry. Next, adding the vectors
$b_{4,5},\alpha$,
the number of generations is reduced to three and the observable-sector 
gauge 
group obtained is $SU(5)\times U(1)$ accompanied by additional four $U(1)$ 
factors and a hidden-sector $SU(4)\times SO(10)$ gauge symmetry.
The massless spectrum includes
the seventy chiral superfields listed in Table 1, together with their
non-Abelian group representation contents and their
$U(1)$ charges~\cite{aehn}.

As seen explicitly in Table 1, the matter and
Higgs fields in this string model carry additional charges under extra
$U(1)$ symmetries~\cite{aehn}, there are a number
of neutral singlet fields, and some hidden-sector 
matter fields which transform
non-trivially under the $SU(4)\times SO(10)$ gauge symmetry:
sextets under $SU(4)$, namely $\Delta_{1,2,3,4,5}$, and
decuplets under $SO(10)$, namely $T_{1,2,3,4,5}$. There are also
fourplets of the $SU(4)$ hidden symmetry which possess fractional
charges. However, these are confined and
will not concern us here.

We recall that the flavour assignments of the light
Standard Model particles in this model are as follows:
\be
\bar{f}_1 : \bar{u}, \tau;  \;
\bar{f}_2 : \bar{c}, e/ \mu;  \;
\bar{f}_5 : \bar{t}, \mu / e; \;
F_2 : Q_2, \bar{s}; \; 
F_3 : Q_1, \bar{d}; \; 
F_4 : Q_3, \bar{b}; \;
\ell^c_1 : \bar{\tau}; \;
\ell^c_2 : \bar{e}; \;
\ell^c_5 : \bar{\mu}
\label{assignments}
\ee
and the trilinear superpotential relevant to fermion and Higgs
doublet or triplet masses is
\begin{eqnarray}
W^3&=&
\frac12 F_1 F_1 h_1+\frac12 F_2 F_2 h_2+\frac12 F_4 F_4 h_1+
\frac12 \bar F_5 \bar F_5 \bar h_2+ F_4\bar f_5\bar h_{45}+
F_3\bar f_3\bar h_3\nonumber\\
&&\mbox{}+\bar f_1 \ell^c_1 h_1+\bar f_2 \ell^c_2 h_2+
\bar f_5 \ell^c_5 h_2+
 h_1\bar h_2 \Phi_{12}+ h_2\bar h_1 \bar\Phi_{12}
+ h_2\bar h_3 \Phi_{23}+ h_3\bar h_2 \bar\Phi_{23}
\nonumber\\ &&\mbox{}
+ h_3\bar h_1 \Phi_{31}+ h_1\bar h_3 \bar\Phi_{31} +h_3\bar h_{45} \bar\phi_{45}
+h_{45}\bar h_3 \phi_{45} +\frac12 h_{45}\bar h_{45}\Phi_3
\end{eqnarray}
As also seen in Table 1, there are four candidate pairs of
electroweak Higgs doublets, and a corresponding number of
partner Higgs triplets in the multiplets $h_1, h_2, h_3$ and $h_{45}$
and $\bar h_1, \bar h_2, \bar h_3$ and $\bar h_{45}$
shown in Table 1. There are also candidates for the GUT Higgs
multiplets among the $F_{1,2,3,4}$ and ${\bar F}_5$ shown in Table 1.

Their tree-level couplings are sufficient to give masses to one
pair of Higgs triplet components. The question we address in this
paper is whether non-renormalizable higher-order terms in the
superpotential can give masses to the remaining Higgs triplets,
while leaving light at least one pair of Higgs doublets.

The string-derived flipped $SU(5)$ model contains an anomalous $U(1)_A$
gauge factor, leading to symmetry breaking via vev's at the scale $M_A\sim
10^{-1} M_{P}$ for GUT-singlet fields that 
are also listed in Table 1. The superpotential of the model
contains, in addition to tree-level trilinear terms,
higher-order non-renormalizable terms. If all but three (two)
of the fields in such an $n^{th}$-order interaction acquire large vev's $V
\sim
10^{-1} M$, one will be left with a residual trilinear (bilinear)
interaction with coefficient of order $10^{3-n} M$.
Such interactions may provide interesting and realistic textures for
quark, charged-lepton and neutrino mass matrices~\cite{ELLN}, since they
provide entries that are hierarchically suppressed. Models of
this type must make choices of the field vev's that are consistent
with the $D$ and $F$ flatness of the effective scalar potential.

It is not an easy task to secure the existence of massless
electroweak doublets in such a model, while also keeping their
triplet partners massive. Consider a generic doublet mass
term of the general form
\be
g \left(\frac{\Phi}{M}\right)^{n - 3} \Phi\,\bar
h\,h
\label{genericmass}
\ee
where $\Phi$ represents a generic field that obtains a large
vev $V \sim {\cal O}(M / 10)$.
Even a 
term of order $n=17$ of the type (\ref{genericmass})
could in
principle push the Higgs masses above the electroweak scale.
In principle, one should calculate all the relevant
non-renormalizable terms up to $17^{\rm th}$ order, in order
to ensure that 
there exist flat directions that preserve at least one
massless pair of Higgs doublets. This paper does not go that
far, but we do extend the analysis to tenth order, as
discussed in following sections.

\section{Conditions for Light Higgs Doublets}

We now discuss how light Higgs doublets may appear,
starting with the tree-level
contributions to the superpotential.
All phenomenologically
viable flat directions require non-zero
vev's for the singlet fields $\Phi_{31},\bar\Phi_{31}$,
$\Phi_{23},\bar\Phi_{23}$ and $\phi_{45},\bar\phi_{45}$
shown in Table 1, which
enter the Higgs mass matrix arising from the
$SU(5)$ representations $h_{1,2,3}$, $h_{45}$ and
$\bar h_{1,2,3}$, $\bar h_{45}$ already at the tree level.
We recall that the tree-level
superpotential terms which may provide the third-generation
masses involve the Higgses
$h_{1,2}$ and  $\bar h_{45}$. The
tree-level doublet Higgs matrix has two 
zero eigenvalues, and one pair of the corresponding eigenvectors 
do indeed have $h_{45}$ and $h_{1,2}$ Higgs fields as components.
The doublet Higgs mass matrix at
the tree level is:
\begin{equation}
M_{2}^3 =  \left(\begin{array}{cccc}
0  & \Phi_{12} & \bar\Phi_{31} & 0  \\
\bar{\Phi}_{12} &   0   & \Phi_{23} &0 \\
\Phi_{31}  &  \bar\Phi_{23}& 0  &\bar\phi_{45} \\
0 &0&\phi_{45}&\Phi_3
\end{array}\right), \label{higgsi}
\end{equation}
but with our choice of singlet vacuum expectation values
$\langle \Phi_{12},\bar{\Phi}_{12}\rangle = \langle \bar{\Phi}_{3}\rangle=0$, so
there are two massless pairs~\cite{KTJR}.

To ensure the masslessness of suitable Higgs mass eigenstates
when higher-order non-renormalizable terms are included, we work as
follows.
We first consider the most general texture for the $4\times 4$ 
doublet Higgs mass
matrix, including arbitrary contributions to the entries that
vanish at the tree level: 
\begin{equation}
M_{2}^{all} =  \bordermatrix{
&\bar h_1&\bar h_2&\bar h_3&\bar h_{45}\cr
h_1&  \ve1& \ve{2} & \bar\Phi_{31} &  \ve3 \cr
h_2&\ve{4} &  \ve5   & \Phi_{23} &\ve6 \cr
h_3&\Phi_{31}  &  \bar\Phi_{23}& \ve7  &\bar\phi_{45} \cr
h_{45}&\ve8 &\ve9&\phi_{45}&\ve{10}\cr}
, \label{higgs}
\end{equation}
where we may ignore higher-order contributions to the 
entries which are non-zero at the tree level.
The parameters $\varepsilon_{1,...,10}$ correspond to all
possible non-renormalizable contributions. 
Next, we check which of  the parameters $\varepsilon_{1,...,10}$
must be zero in order to obtain zero eigenvalues whose
eigenvectors have components along the $h_{1,2}$ and $\bar{h}_{45}$
directions. We find 54 solutions.
Whether $\varepsilon_7 = 0$ or $\not= 0$ does not affect the presence
of such massless eigenstates. Factoring out these two options,
there remain 27 options for combinations of the other
$\varepsilon_i$ that need not vanish, as listed in Tables 2 and \ref{oht}.

The next task is to analyze the Higgs
triplet mass matrix, which takes the form
\begin{equation}
M_{3}^{all} =  \bordermatrix{
&\bar h_1&\bar h_2&\bar h_3&\bar h_{45}&F_1\cr
h_1&  \ve1& \ve{2} & \bar\Phi_{31} &  \ve3&\lambda_1 \cr
h_2&\ve{4} &  \ve5   & \Phi_{23} &\ve6 &\lambda_2\cr
h_3&\Phi_{31}  &  \bar\Phi_{23}& \ve7  &\bar\phi_{45}&\lambda_3 \cr
h_{45}&\ve8 &\ve9&\phi_{45}&\ve{10}&\lambda_4\cr
{\bar F}_5&\bar\lambda_1&\bar\lambda_2&\bar\lambda_3&\bar\lambda_4&0\cr}
, \label{higgst}
\end{equation}
where the $\lambda_i$ and ${\bar \lambda}_j$: $i,j = 1,2,3,4$
are generic Yukawa couplings. Two of these, namely
$\lambda_1$ and ${\bar \lambda}_2$, are non-vanishing at the
tree level. The issue then is: for each of the doublet options listed
in Tables 2 and 3, which combinations of the $\lambda_i$ and ${\bar
\lambda}_j$ must be non-vanishing in order that all the Higgs triplets
are massive?

The answers are that no such satisfactory combinations exist
for the two Higgs doublet textures listed in Table 2, whereas
there are possible solutions for each of the other 52 textures,
as shown in the last column of Table 3.

\section{Non-Renormalizable Contributions to the Higgs Mass Textures}

We have enumerated all possible non-renormalizable terms up to
tenth order, but do not worry: we shall not list them all here.
Instead, we present only the contributions up to seventh order in
Table 4, and then we discuss  those higher-order terms that have a chance
of being
non-vanishing. For this purpose, we take into account the
choices of non-zero field vev's that have been made in previous
phenomenological studies of the string flipped $SU(5)$ model.
We therefore assume that just the following
minimal set of fields have vanishing vev's, satisfying all flatness conditions
\cite{ART,ELLN}
\bea
&&\Ph12=\Phb12=\Phi_{I=1,\dots,5}=\phi_1=\phi_3=\bar\phi_1=\bar\phi_2=0\nonumber\\
&&\phi^{+}=\bar\phi^{-}=\bar\phi_3 \phi_4=F_4=F_3=F_2=\Delta_1=T_1=0,
\label{zeroes}
\eea
and that the following set of non-Abelian composite gauge-singlet
condensates also vanish~\footnote{Here we
introduce
the notation $\left[A_1\dots A_n\right]$ to denote a general
linear combination of 
all possible group invariants of the fields $A_1,A_2,\dots A_n$.}
:
\be
\si{T_4 T_4}=\si{T_5 T_5}=\si{\Delta_4 \Delta_4}=\si{\Delta_4\Delta_5}=
\si{\Delta_5 \Delta_5}=
\si{T_2 T_4}=\si{T_4 T_5}=\si{\D2\D2}+\si{T_2 T_2}=0,
\label{compzeroes}
\ee
We also assume that the following minimal set 
of elementary fields have non-vanishing vev's
\be
\Ph23,\Phb23,\Ph31,\Phb31,\ph45,\phb45,\phi_2,\bar\phi_4,
F_1,\bar F_5 \not= 0
\label{nonzeroes}
\ee
as well as the following composite fields:
\be
\si{\Delta_2\Delta_3} \not= 0
\  \mbox{and}\ \ \ 
\left\{\begin{array}{l}
 \si{T_3 T_4}\not=0\\ \mbox{or}\\ \si{T_3 T_5}\not=0
\end{array}\right. .
\label{compnonzeroes}
\ee
We
make no {\it a priori} assumption about the vev's of the remaining 
elementary or composite fields.

With the above choice (\ref{zeroes},\ref{compzeroes},\ref{nonzeroes},\ref{compnonzeroes}) of vev's, 
the first non--renormalizable
terms in the
doublet mass matrix appear at seventh order. This means that the $\ve{i}$
are
seventh order or higher, so that the Higgs doublet masses are certainly
much smaller than $M_{GUT}$.
In order to have the triplet masses as heavy as possible, we search for
solutions involving
only the tree--level contributions to the
triplet couplings $\lambda_i$,$\bar\lambda_i$.
Since the tree--level superpotential gives $\l1=\langle F_1\rangle,
\lb2=\langle\bar F_5\rangle$, we have
$\lambda_{i\not=1}=\bar\lambda_{i\not=2}=0$ to this order.
Examining the last column of Table 3 for textures leading to
non--zero triplet masses, we find the
14 options (3-7,10-15,22-24), for each of which $\ve7 \not= 0$
is a possible option.

All these textures need $\ve3=\ve2=0$, so we must check these
conditions as far as possible. Imposing
$\ve3=0$ up to ninth order yields the conditions
\begin{eqnarray}
&&\si{T_2 T_3}=\si{\D3 \D4}=\si{\D3\D5}\bar\phi_3=\si{T_3 T_4 T_3 T_4}\si{\D3 \D3}=0
\nonumber\\
&&\si{T_2 T_5} \si{\D2\D5}= \si{T_2 T_5} \si{\D2\D4}\bar\phi_3=0\nonumber\\
&&\bar\phi_{+}\si{\D3\D3\D5\D5}=\bar\phi_{+}\si{\D3\D3\D4\D4}=
\bar\phi_{3}\si{\D3\D3\D5\D4}=0\label{ep3}\\
&&\phi_{-}\left(\si{\D3\D3\D2\D2}+\si{\D3\D3}\si{T_2 T_2}\right)=
\bar\phi_{+}\left(\si{\D3\D3\D2\D2}+\si{\D3\D3}\si{T_2 T_2}\right)=0\nonumber\\
&&\si{\D5\D5\D2\D2\D2\D2}+\si{\D5\D5\D2\D2 T_2 T_2}=0\nonumber
\end{eqnarray}
Further imposing
$\ve2=0$ up to ninth order yields the additional conditions
\begin{eqnarray}
&&\si{T_4 T_4 T_4 T_5 T_5 T_5}=0\\
&&\si{T_5 T_4 T_4 T_4}\si{\D2\D2}+\si{T_5 T_4 T_4 T_4 T_2 T_2}=0
\end{eqnarray}
Both sets of these conditions are simultaneously compatible with
our initial choice (\ref{zeroes},\ref{compzeroes},\ref{nonzeroes},\ref{compnonzeroes})
of vev's.

Next we search for the lowest--order non--vanishing $\ve{i}$. 
We find none at seventh order, and to
eighth order only $\ve1,\ve4,\ve6,\ve{10}\not=0$. We note also that $\ve1\propto\ve{10}$ and $\ve4\propto\ve6$.
Among these options,
only $\ve1=\ve{10}=0$ together with $\ve6,\ve4\not=0$ give acceptable textures.
The condition
$\ve1=\ve{10}=0$ then imposes the extra constraint
\begin{equation}
\si{\D5\D4\D2\D2}=0.
\label{ve10vanish}
\end{equation}
This selects the textures 15, and possibly 13 if ${\ve5 }$ is
generated at higher order. 

All of $\ve{1,8,9,10}$ have to vanish in order to
preserve one massless doublet pair. 
Examining the ninth-order Higgs mass terms, we find that
$\ve5=\ve9=\ve{10}=0$ automatically, whilst
in order to keep $\ve1=\ve8=0$ we have to impose
\begin{eqnarray}
&&\si{\D5\D5\D4\D4\D2\D2}+ \si{\D5\D5\D4\D4}\si{T_2 T_2}= 0\\
&&\si{\D5\D5\D5\D4\D2\D2}+ \si{\D5\D5\D5\D4}\si{T_2 T_2}= 0\\
&&\si{\D5\D5\D5\D5\D2\D2}+ \si{\D5\D5\D5\D5}\si{T_2 T_2}= 0
\label{nove5}
\end{eqnarray}
Again, these conditions are compatible with the previous
choices  of vacuum parameters, and amount to a refinement of
the string vacuum choice.

With the above vacuum choice, the relevant non--renormalizable
contributions to the 
doublet mass matrix up to the ninth order are
\begin{eqnarray}
\ve4&=&{\cal O}(1)\frac{1}{M^5}\left(\si{\D5\D5\D3\D3}\phi_{-}\phi_{45}\right)\\
\ve6&=&{\cal O}(1)\frac{1}{M^5}\left(\si{\D5\D5\D3\D3}\phi_{-}
\bar\Phi_{31}\right)
\end{eqnarray}
In order to insure a non--vanishing $\ve6,\ve4$, we have to impose
\begin{equation}
\si{\D5\D5\D3\D3},\phi_{-}\not=0
\label{ve4not0}
\end{equation}
When combined with conditions (\ref{ep3}), this implies
that the vacuum must also have
\begin{equation}
\bar\phi_{+}=\si{\D2\D2\D3\D3}+\si{\D2\D2}\si{\D3\D3}=0
\label{vaccondn}
\end{equation}
which completes our specification of the string vacuum
to this order.

We discuss now the Higgs triplet masses. Analyzing texture 15, we first
note that there are four pairs which are massive at the
tree level, two of them with masses proportional to $\l1\lb2$.
The fifth pair has the lightest mass, which is
\begin{equation}
M_3^{light}\sim\frac{|\ph45|\left|\ve4\phb45-\ve6\Ph31\right|}{(\sqrt{|\Ph23|^2+|\ph45|^2)
(|\Ph31|^2+|\phb45|^2)}}
\label{tripletmass}
\end{equation}
for our vacuum choice. If we assume generic vev's ${\cal O}(1/10)$
in natural units, we see that (\ref{tripletmass}) indicates that
the lightest triplet mass might be in the range of $10^{10}$ to
$10^{12}$~GeV, which is perfectly satisfactory in flipped $SU(5)$.

We now turn to the Higgs doublets. By construction, there is a massless
pair, and there were two massive pairs {\it ab initio} at the tree level.
The fourth Higgs pair is relatively light, with mass
\begin{equation}
M_2^{light}\sim\frac{\left(|\ph45|^2+|\Phb31|^2\right)^{\frac12}\left(
\left(|\ve4|^2+|\ve6|^2\right)|\Phb23|^2+\left|\ve6\Ph31-\ve4\phb45\right|^2\right)^\frac12}
{\left(|\Ph23|^2+|\ph45|^2+|\Phb31|^2\right)^{\frac12}\left(
|\Ph31|^2+|\Phb23|^2+|\phb45|^2\right)^{\frac12}}
\label{doubletmass}
\end{equation}
It is interesting that this mass is comparable in order of magnitude
to the lightest triplet mass (\ref{tripletmass}), whereas all the other
massive doublets and triplets have masses comparable to $M_{GUT}$.
Thus, below $M_{GUT}$, we have (effectively) only complete GUT
representations, and the standard supersymmetric GUT prediction for
the electroweak mixing angle is at least not affected to first order by
their appearance in the renormalization-group equations.

According to Table 3, the Higgs doublets that are massless
to ninth order are the combinations
\begin{eqnarray}
h_0&=& \frac{1}{\sqrt{\phb45^{*2}+\Phb31^{*2}}}\left(
\phb45^* h_1-\Phb31^{*} h_{45}\right)
\label{hzero}\\
{\bar h}_0&=&
\frac{1}{\sqrt{\Phb23^2(1+r^2) +(\ph45-\Ph31 r)^2}}
\left(-\Phb23 r \bar h_1
+(\ph45-\Ph31 r)\bar h_2 -\Phb23 \bar h_{45}\right)
\label{hbarzero}
\end{eqnarray}
Where $r=\frac{\ve6}{\ve4}={\cal O}(1)\frac{\ph45}{\Phb31}$.
Moreover, we have checked that this light Higgs doublet pair
remains massless when tenth-order terms
 in the superpotential
 involving observable field vev's
are taken into account.

We note that $h_0$ (\ref{hzero}) contains components with tree-level
couplings
only to the 
third-generation quark and lepton
masses. This is welcome since the two lighter generations receive masses 
from non--renormalizable terms:
up to fifth order one finds
\begin{equation}
W_5= F_2\bar f_2\bar h_{45} + \bar F_2 \bar f_2 \bar h_{45}\bar\phi_4+
F_2 F_2 h_1\left(\bar\phi_4^2+\bar\phi_3^2\right)
+\bar f_2 \ell^c_2 h_1\left(\bar\phi_4^2+\bar\phi_3^2\right)+\dots.
\label{superpot5}
\end{equation}
Thus, a natural hierarchy between the third- and second-generation
quark and lepton masses is a prediction of our doublet-triplet splitting
mechanism.

Let us now comment on the uniqueness of the solution
presented above. It 
is the only one that involves tree--level couplings in both the $\l{i}$
and the
$\lb{i}$. If one relaxes this constraint,
which will tend to reduce the lightest triplet Higgs mass, many other
textures
in Table 3 can provide
acceptable solutions. For example, a search  for
$FF h$ and $\bar F\bar F \bar h$ couplings to fifth order yields
already
\begin{eqnarray}
\l2&\sim&F_1(\phi_2^2+\phi_4^2)\\
\l4&\sim&F_1\ph45\Ph31\\
\lb1&\sim&\bar F_{5}(\phi_2^2+\phi_4^2)\\
\lb4&\sim&F_1\phb45\Ph23.
\end{eqnarray}
As can easily be checked from Table 3,  $\l3\lb3$ is irrelevant and we
have not calculated it.

\section{Conclusions and Prospects}

We have presented in this paper a solution to the triplet--doublet
splitting problem in which
the light Higgs doublets do not acquire a mass 
from any superpotential term through ninth order, nor even at
tenth order if we discard vev's for higher-order combinations
of hidden-sector fields. This solution has the attractive features
that there are effectively complete GUT representations of
intermediate-mass Higgs doublets and triplets, and predicts
that the second-generation charge -1/3 quarks and leptons are
much lighter than those in the third generation.
As explained earlier, one really needs to check this solution up to
$17^{th}$ order in the superpotential in order to ensure that the light
Higgs doublet is sufficiently light. However, the analysis
presented here goes to much higher order than had been done
previously, and constitutes a promising start. If, eventually,
this solution does not survive, there are other possible solutions,
as mentioned at the end of the previous section.
We are optimistic that the version of flipped $SU(5)$ derived
from string can live up to its field-theoretical promise of
solving the doublet-triplet mass splitting problem.

\vspace*{2cm}
\noindent
{\bf Acknowledgements}
~~\\
\noindent
The work of J.R. was supported in part by the EU under the TMR
contract ERBFMRX-CT96-0090.

\newpage
\begin{table}[h]
\centering
\begin{tabular}{lll}
\hline
$F_1({\bf{10}},\frac{1}{2},-\frac{1}{2},0,0,0)$ &
$\bar{f}_1({\bf\bar5}-\frac{3}{2},-\frac{1}{2},0,0,0)$ &
$\ell_1^c({\bf1},\frac{5}{2},-\frac{1}{2},0,0,0)$ \\
$F_2({\bf10},\frac{1}{2},0,-\frac{1}{2},0,0)$ &
$\bar{f}_2({\bf\bar5}-\frac{3}{2},0,-\frac{1}{2},0,0)$ &
$\ell_2^c({\bf1},\frac{5}{2},0,-\frac{1}{2},0,0)$ \\
$F_3({\bf10},\frac{1}{2},0,0,\frac{1}{2},-\frac{1}{2})$ &
$\bar{f}_3({\bf\bar5}-\frac{3}{2},0,0,\frac{1}{2},\frac{1}{2})$ &
$\ell_3^c({\bf1},\frac{5}{2},0,0,\frac{1}{2},\frac{1}{2})$ \\
$F_4({\bf10},\frac{1}{2},-\frac{1}{2},0,0,0)$ &
$f_4({\bf5},\frac{3}{2},\frac{1}{2},0,0,0)$ &
$\bar\ell_4^c({\bf1},-\frac{5}{2},\frac{1}{2},0,0,0)$ \\
$\bar{F}_5({\bf\overline{10}},-\frac{1}{2},0,\frac{1}{2},0,0)$ &
$\bar{f}_5({\bf\bar5}-\frac{3}{2},0,-\frac{1}{2},0,0)$ &
$\ell_5^c({\bf1},\frac{5}{2},0,-\frac{1}{2},0,0)$ \\
\hline
\\
\hline
$h_1({\bf5},-1,1,0,0,0)$ & $h_2({\bf5},-1,0,1,0,0)$ & $h_3({\bf5},-1,0,0,1,0)$ \\
$h_{45}({\bf5},-1,-\frac{1}{2},-\frac{1}{2},0,0)$ & & \\
$\bar h_1({\bf\bar5},1,-1,0,0,0)$ & $\bar h_2({\bf\bar5},1,0,-1,0,0)$ 
& $\bar h_3({\bf\bar5},1,0,0,-1,0)$ \\
$\bar h_{45}({\bf\bar5},1,\frac{1}{2},\frac{1}{2},0,0)$ & & \\
\hline
\\
\hline
$\phi_{45}({\bf1},0,\frac{1}{2},\frac{1}{2},1,0) $ &
$\phi_{+}({\bf1},0,\frac{1}{2},-\frac{1}{2},0,1) $ &
$\phi_{-}({\bf1},0,\frac{1}{2},-\frac{1}{2},0,-1) $ \\
$\bar\phi_{45}({\bf1},0,-\frac{1}{2},-\frac{1}{2},-1,0) $ &
$\bar\phi_{+}({\bf1},0,-\frac{1}{2},\frac{1}{2},0,-1) $ &
$\bar\phi_{-}({\bf1},0,-\frac{1}{2},\frac{1}{2},0,1) $ \\
$\Phi_{23}({\bf1},0,0,-1,1,0)$ &
$\Phi_{31}({\bf1},0,1,0,-1,0)$  &
$\Phi_{12}({\bf1},0,-1,1,0,0)$ \\
$\phi_i({\bf1},0,\frac{1}{2}, -\frac{1}{2},0,0)$ &
$\bar\phi_i({\bf1},0,-\frac{1}{2}, +\frac{1}{2},0,0)$,&i=1,2,3,4 \\
$\bar\Phi_{23}({\bf1},0,0,1,-1,0)$ &
$\bar\Phi_{31}({\bf1},0,-1,0,1,0)$  &
$\bar\Phi_{12}({\bf1},0,1,-1,0,0)$ \\
$\Phi_i({\bf1},0,0,0,0,0)$, &I=1,2,3,4,5  &\\
\hline
\\
\hline
$\Delta_1(0,{\bf1},{\bf{6}},0,-\frac{1}{2},\frac{1}{2},0)$ &
$\Delta_2(0,{\bf1},{\bf{6}},-\frac{1}{2},0,\frac{1}{2},0)$ &
$\Delta_3(0,{\bf1},{\bf{6}},-\frac{1}{2},-\frac{1}{2},0,
\frac{1}{2})$ \\
$\Delta_4(0,{\bf1},{\bf{6}},0,-\frac{1}{2},\frac{1}{2},0)$ &
$\Delta_5(0,{\bf1},{\bf{6}},\frac{1}{2},0,-\frac{1}{2},0)$ & \\
$T_1(0,{\bf10},{\bf1},0,-\frac{1}{2},\frac{1}{2},0)$ &
$T_2(0,{\bf10},{\bf1},-\frac{1}{2},0,\frac{1}{2},0)$ &
$T_3(0,{\bf10},{\bf1},-\frac{1}{2},-\frac{1}{2},0,\frac{1}{2})$ \\
$T_4(0,{\bf10},{\bf1},0,\frac{1}{2},-\frac{1}{2},0)$ &
$T_5(0,{\bf10},{\bf1},-\frac{1}{2},0,\frac{1}{2},0)$ & \\
\hline
\end{tabular}
\caption{
{\it The chiral superfields of the version of the flipped $SU(5)$
model derived from string, with their quantum numbers~\cite{aehn}.
For $F_i$, $\bar{f}_i$, $\ell_i^c$,
$h_i$, $\bar h_{i}$  and the singlet fields
the 
$ SU(5) \times U(1)' \times U(1)^4$ quantum numbers are presented.
For 
 $\Delta_i$ and $T_i$ we present the 
 $U(1)' \times SO(10) \times SO(6) \times U(1)^4$
quantum numbers. }
}
\end{table}
\begin{table}
\begin{center}
\begin{tabular}{|l|c|c|c|}
\hline
&$M_2$&$h_0, h_0'$&$\bar h_0,\bar h_0'$\\
\hline
1&
$
\left(
\begin{array}{cccc}
0     &0      &\Phb31 &0      \\
0     &0      &\Ph23  &0      \\
\Ph31 &\Phb23 &0/\ve7      &\phb45 \\
0     &0      &\ph45  &0      \\
\end{array}
\right)
$
&
$\left[\begin{array}{c}
\vphantom{-}\cph45\\
0\\
0\\
-\cPhb31\\
\end{array}
\right]
$
\ ,\ 
$\left[\begin{array}{c}
\cPhb23\\
-\cPhb31\\
0\\
0\\
\end{array}
\right]
$
&
$\left[\begin{array}{c}
\cph45\\
0\\
0\\
-\cPhb31\\
\end{array}
\right]
$
\ ,\ 
$\left[\begin{array}{c}
\cPhb23\\
-\cPhb31\\
0\\
0\\
\end{array}
\right]
$\\
\hline
\end{tabular}
\end{center}
\caption{Textures leading to two pairs of massless Higgs doublets, with
unnormalized indications of the massless
eigenstates ($h_0,h_0'$, $\bar h_0, \bar h_0'$).}
\end{table}
 
%$$ \hm{0}{0}{\Phb31}{0}{0}{0}{\Ph23}{0}{\Ph31}{\Phb23}{0}{\phb45}{0}{0}{\ph45}{0}$$
\begin{table}
\begin{center}
{\small
\begin{tabular}{|l|c|c|c|c|}
\hline
&$M_2$&$h_0$&$\bar h_0$&$\det M_3$\\
\hline
2&
$
\left(
\begin{array}{cccc}
0     &0      &\Phb31 &0      \\
0     &0      &\Ph23  &0      \\
\Ph31 &\Phb23 &0/\ve7      &\phb45 \\
0     &\ve9   &\ph45  &0      \\
\end{array}
\right)
$
&
$\left[\begin{array}{c}
\phantom{-}\cPh23\\
-\cPhb31\\
\phantom{-}0\\
\phantom{-}0\\
\end{array}
\right]
$
&
$\left[\begin{array}{c}
\phb45\\
\phantom{-}0\\
\phantom{-}0\\
-\Ph31\\
\end{array}
\right]
$
&$\ve9(\l1-\l2)(\lb1-\lb4)$
\\ \hline
3&
$
\left(
\begin{array}{cccc}
0     &0      &\Phb31 &0      \\
0     &0      &\Ph23  &0      \\
\Ph31 &\Phb23 &0/\ve7      &\phb45 \\
0     &\ve9   &\ph45  &\ve{10}    \\
\end{array}
\right)
$
&
$\left[\begin{array}{c}
\phantom{-}\cPh23\\
-\cPhb31\\
\phantom{-}0\\
\phantom{-}0\\
\end{array}
\right]
$
&
$\left[\begin{array}{c}
\Phb23\ve{10}-\phb45\ve9\\
-\Ph31\ve{10}\\
0\\
\Ph31\ve9\\
\end{array}
\right]
$
&\begin{minipage}{4cm}$
(\Ph23\l1-\Phb31\l2)$\\ $\times(\phb45\ve{9}\lb1- \Phb23\ve{10}\lb1$\\$ + \Ph31\ve{10}\lb2 - \Ph31\ve9\lb4)
$\end{minipage}
\\ \hline
4&
$
\left(
\begin{array}{cccc}
0     &0      &\Phb31 &0      \\
0     &0      &\Ph23  &0      \\
\Ph31 &\Phb23 &0/\ve7      &\phb45 \\
\ve8  &0      &\ph45  &0      \\
\end{array}
\right)
$
&
$\left[\begin{array}{c}
\phantom{-}\cPh23\\
-\cPhb31\\
\phantom{-}0\\
\phantom{-}0\\
\end{array}
\right]
$
&
$\left[\begin{array}{c}
\phantom{-}0\\
\phantom{-}\phb45\\
\phantom{-}0\\
-\Phb23\\
\end{array}
\right]
$
&\begin{minipage}{4cm}$\ve8 (\Ph23\l1-\Phb31\l2)$\\ $\times(\phb45\lb2-\Phb23\lb4)$\end{minipage}
\\\hline
5&
$
\left(
\begin{array}{cccc}
0     &0      &\Phb31 &0      \\
0     &0      &\Ph23  &0      \\
\Ph31 &\Phb23 &0/\ve7      &\phb45 \\
\ve8     &0   &\ph45  &\ve{10}      \\
\end{array}
\right)
$
&
$\left[\begin{array}{c}
\phantom{-}\cPh23\\
-\cPhb31\\
\phantom{-}0\\
\phantom{-}0\\
\end{array}
\right]
$
&
$\left[\begin{array}{c}
\Phb23\ve{10}\\
\phb45\ve8-\Ph31\ve{10}\\
0\\
-\Phb23\ve8\\
\end{array}
\right]
$
&
\begin{minipage}{4cm}$
(\Ph23\l1-\Phb31\l2)$\\$\times (\phb45\ve8\lb2- 
\Ph31\ve{10}\lb2$\\$ + \Phb23 \ve{10}\lb1 - \Phb23\ve8\lb4)$
\end{minipage}
\\ \hline
6&
$
\left(
\begin{array}{cccc}
0     &0      &\Phb31 &0      \\
0     &0      &\Ph23  &0      \\
\Ph31 &\Phb23 &0/\ve7      &\phb45 \\
\ve8  &\ve9   &\ph45  &0      \\
\end{array}
\right)
$
&
$\left[\begin{array}{c}
\phantom{-}\cPh23\\
-\cPhb31\\
\phantom{-}0\\
\phantom{-}0\\
\end{array}
\right]
$
& 
$\left[\begin{array}{c}
\phb45\ve9\\
-\phb45\ve8\\
0\\
\Phb23\ve8-\Ph31\ve9\\
\end{array}
\right]
$
&
\begin{minipage}{4cm}$
-(\Ph23\l1-\Phb31\l2)$\\$\times(\phb45 \ve9\lb1- \phb45\ve8\lb2$\\$ + \Phb23\ve8\lb4 - \Ph31\ve9\lb4)$
\end{minipage}
\\\hline
7&
$
\left(
\begin{array}{cccc}
0     &0      &\Phb31 &0      \\
0     &0      &\Ph23  &0      \\
\Ph31 &\Phb23 &0/\ve7      &\phb45 \\
\ve8  &\ve9   &\ph45  &\ve{10}    \\
\end{array}
\right)
$
&
$\left[\begin{array}{c}
\phantom{-}\cPh23\\
-\cPhb31\\
\phantom{-}0\\
\phantom{-}0\\
\end{array}
\right]
$
& 
$\left[\begin{array}{c}
\Phb23\ve{10}-\phb45\ve9\\
\phb45\ve8-\Ph31\ve{10}\\
0\\
\Ph31\ve9-\Phb23\ve8\\
\end{array}
\right]
$
&
\begin{minipage}{4cm}
$-(\Ph23\l1-\Phb31\l2)$\\$\times(\phb45\ve9\lb1- \Phb23\ve{10}\lb1$\\$ - \phb45\ve8\lb2 + \Ph31\ve{10}\lb2+
\Phb23\ve8\lb4-\Phb31\ve9\lb4)$
\end{minipage}
\\\hline
8&
$
\left(
\begin{array}{cccc}
0     &0      &\Phb31 &0      \\
0     &\ve5   &\Ph23  &0      \\
\Ph31 &\Phb23 &0/\ve7      &\phb45 \\
0     &0      &\ph45  &0      \\
\end{array}
\right)
$
&
$\left[\begin{array}{c}
\phantom{-}\cph45\\
\phantom{-}0\\
\phantom{-}0\\
-\cPhb31\\
\end{array}
\right]
$
& 
$\left[\begin{array}{c}
\phantom{-}\phb45\\
0\\
0\\
-\Ph31
\end{array}
\right]
$
&\begin{minipage}{4cm}$\ve5 (\ph45\l1-\Phb31\l4)$\\$\times(\phb45\lb1-\Ph31\lb4)$\end{minipage}
\\\hline
9&
$
\left(
\begin{array}{cccc}
0     &0      &\Phb31 &0      \\
0     &\ve5   &\Ph23  &0      \\
\Ph31 &\Phb23 &0/\ve7      &\phb45 \\
0     &\ve9   &\ph45  &0      \\
\end{array}
\right)
$
&
$\left[\begin{array}{c}
\cPh23\cve9-\cph45\cve5\\
-\cPhb31\cve9\\
0\\
\cPhb31\cve5\\
\end{array}
\right]
$
& 
$\left[\begin{array}{c}
\phantom{-}\phb45\\
0\\
0\\
-\Ph31\\
\end{array}
\right]
$
&
\begin{minipage}{4cm}
$
(\phb45\lb1 - \Ph31\lb4)$\\$\times(\ph45\ve5\l1- \Ph23\ve9\l1 $\\$+ \Phb31\ve9\l2 - \Phb31\ve5\l4)
$\end{minipage}
\\\hline
10&
$
\left(
\begin{array}{cccc}
0     &0      &\Phb31 &0      \\
0     &\ve5   &\Ph23  &\ve6      \\
\Ph31 &\Phb23 &0/\ve7      &\phb45 \\
0     &0      &\ph45  &0      \\
\end{array}
\right)
$
&
$\left[\begin{array}{c}
\phantom{-}\cph45\\
\phantom{-}0\\
\phantom{-}0\\
-\cPhb31\\
\end{array}
\right]
$
& 
$\left[\begin{array}{c}
\Phb23\ve6-\phb45\ve5\\
-\Ph31\ve6\\
0\\
\Ph31\ve5\\
\end{array}
\right]
$
&
\begin{minipage}{4cm}
$(\ph45\l1 - \Phb31\l4)$\\$\times(\phb45\ve5\lb1- \Phb23\ve6\lb1$\\$ + \Ph31\ve6\lb2 - \Ph31\ve5\lb4)$
\end{minipage}
\\\hline
\end{tabular}
}
\end{center}
\caption{\label{oht}{\it Textures leading to a single pair of massless
Higgs doublets, together with unnormalized indications of the associated
massless
eigenstates ($h_0,\bar h_0$) and the determinant $\det(M_3)$
of the colour--triplet mass matrix.}}
\end{table}
\begin{table}
\begin{center}
\begin{tabular}{|l|c|c|c|c|}
\hline
11&
$
\left(
\begin{array}{cccc}
0     &0      &\Phb31 &0      \\
\ve4  &0      &\Ph23  &0      \\
\Ph31 &\Phb23 &0/\ve7      &\phb45 \\
0     &0      &\ph45  &0      \\
\end{array}
\right)
$
&
$\left[\begin{array}{c}
\phantom{-}\cph45\\
\phantom{-}0\\
\phantom{-}0\\
-\cPhb31\\
\end{array}
\right]
$
& 
$\left[\begin{array}{c}
0\\
\phantom{-}\phb45\\
0\\
-\Phb23\\
\end{array}
\right]
$
%%%%%%%%%
&
\begin{minipage}{4cm}
$\ve4 (\Phb31\l4-\ph45\l1)$\\$ \times (\phb45\lb2-\Phb23 \lb4)$
\end{minipage}
\\\hline
12&
$
\left(
\begin{array}{cccc}
0     &0      &\Phb31 &0      \\
\ve4  &0      &\Ph23  &0      \\
\Ph31 &\Phb23 &0/\ve7      &\phb45 \\
\ve8  &0      &\ph45  &0      \\
\end{array}
\right)
$
&
$\left[\begin{array}{c}
\cPh23\cve8-\cph45\cve4\\
-\cPhb31\cve8\\
\phantom{-}0\\
\cPhb31\cve4\\
\end{array}
\right]
$
& 
$\left[\begin{array}{c}
0\\
\phantom{-}\phb45\\
0\\
-\Phb23\\
\end{array}
\right]
$
&
\begin{minipage}{4cm}
$
-(\phb45\lb2-\Phb23\lb4)$\\
$\times (\ph45 \ve4\l1-
\Ph23\ve8\l1$\\
$+\Phb31\ve8\l2-\Phb31\ve4\l4)$
\end{minipage}
\\\hline
13&
$
\left(
\begin{array}{cccc}
0     &0      &\Phb31 &0      \\
\ve4  &0   &\Ph23  &\ve6      \\
\Ph31 &\Phb23 &0/\ve7      &\phb45 \\
0     &0      &\ph45  &0      \\
\end{array}
\right)
$
&
$\left[\begin{array}{c}
\phantom{-}\cph45\\
\phantom{-}0\\
\phantom{-}0\\
-\cPhb31\\
\end{array}
\right]
$
& 
$\left[\begin{array}{c}
\Phb23\ve6\\
\phb45\ve4-\Ph31\ve6\\
0\\
-\Phb23\ve4\\
\end{array}
\right]
$
&
\begin{minipage}{4cm}
$
(\Phb31\l4-\ph45\l1)$\\$\times (\phb45\ve4\lb2-
  \Ph31\ve6\lb2$\\$+\Phb23\ve6\lb1-\Phb23\ve4\lb4)
$
\end{minipage}
\\\hline
14&
$
\left(
\begin{array}{cccc}
0     &0      &\Phb31 &0      \\
\ve4  &\ve5   &\Ph23  &0      \\
\Ph31 &\Phb23 &0/\ve7      &\phb45 \\
0     &0      &\ph45  &0      \\
\end{array}
\right)
$
&
$\left[\begin{array}{c}
\phantom{-}\cph45\\
\phantom{-}0\\
\phantom{-}0\\
-\cPhb31\\
\end{array}
\right]
$
& 
$\left[\begin{array}{c}
\phb45\ve5\\
-\phb45\ve4\\
0\\
\Phb23\ve4-\Ph31\ve5\\
\end{array}
\right]
$
&
\begin{minipage}{4cm}$
(\ph45\l1-\Phb31\l4)$\\$\times (\phb45\ve5\lb1-
\phb45\ve4\lb2$\\$+\Phb23\ve4\lb4-\Ph31\ve5\lb4)
$\end{minipage}
\\\hline
15&
$
\left(
\begin{array}{cccc}
0     &0      &\Phb31 &0      \\
\ve4  &\ve5   &\Ph23  &\ve6    \\
\Ph31 &\Phb23 &0/\ve7      &\phb45 \\
0     &0      &\ph45  &0      \\
\end{array}
\right)
$
&
$\left[\begin{array}{c}
\phantom{-}\cph45\\
\phantom{-}0\\
\phantom{-}0\\
-\cPhb31\\
\end{array}
\right]
$
& 
$\left[\begin{array}{c}
\phb45\ve5-\Phb23\ve6\\
\phb45\ve4-\Ph31\ve6\\
0\\
\Ph31\ve5-\Phb23\ve4
\end{array}
\right]
$
&
\begin{minipage}{4cm}
$
(\ph45\l1-\Phb31\l4)$\\$\times (\phb45 \ve5 \lb1-
\Phb23\ve6\lb1$\\$-\phb45\ve4\lb2+
\Ph31\ve6\lb2$\\$+\Phb23\ve4\lb4-\Ph31\ve5\lb4)
$\end{minipage}
\\\hline
16&
$
\left(
\begin{array}{cccc}
0     &\ve2   &\Phb31 &0      \\
0     &0      &\Ph23  &0      \\
\Ph31 &\Phb23 &0/\ve7      &\phb45 \\
0     &0      &\ph45  &0      \\
\end{array}
\right)
$
&
$\left[\begin{array}{c}
\phantom{-}0\\
\phantom{-}\cph45\\
\phantom{-}0\\
-\cPh23\\
\end{array}
\right]
$
& 
$\left[\begin{array}{c}
\phantom{-}\phb45\\
0\\
0\\
-\Ph31
\end{array}
\right]
$
&
\begin{minipage}{4cm}
$\ve2 (\Ph23\l4-\ph45\l2)$\\$\times
 (\phb45 \lb1-\Ph31\lb4)$
\end{minipage}
\\\hline
17&
$
\left(
\begin{array}{cccc}
0     &\ve2   &\Phb31 &0      \\
0     &0      &\Ph23  &0      \\
\Ph31 &\Phb23 &0/\ve7      &\phb45 \\
0     &\ve9   &\ph45  &0      \\
\end{array}
\right)
$
&
$\left[\begin{array}{c}
\cPh23\cve9\\
\cph45\cve2-\cPhb31\cve9\\
0\\
-\cPh23\cve2\\
\end{array}
\right]
$
& 
$\left[\begin{array}{c}
\phantom{-}\phb45\\
0\\
0\\
-\Ph31
\end{array}
\right]
$
&
\begin{minipage}{4cm}
$
(\Ph31\lb4-\phb45\lb1)$\\$\times
 (\ph45\ve2\l2-
\Phb31\ve9\l2$\\$
+\Ph23\ve9\l1-\Ph23\ve2\l4)
$\end{minipage}
\\\hline
18&
$
\left(
\begin{array}{cccc}
0     &\ve2   &\Phb31 &0      \\
0     &\ve5   &\Ph23  &0      \\
\Ph31 &\Phb23 &0/\ve7      &\phb45 \\
0     &0      &\ph45  &0      \\
\end{array}
\right)
$
&
$\left[\begin{array}{c}
\cph45\cve5\\
-\cph45\cve2\\
0\\
\cPh23\cve2-\cPhb31\cve5\\
\end{array}
\right]
$
& 
$\left[\begin{array}{c}
\phantom{-}\phb45\\
0\\
0\\
-\Ph31\\
\end{array}
\right]
$
&
\begin{minipage}{4cm}
$
(\phb45\lb1-\Ph31\lb4)$\\$\times
(\ph45\ve5\l1-
\ph45\ve2\l2$\\$
+\Ph23\ve2\l4-\Phb31\ve5\l4)
$\end{minipage}
\\\hline
19&
$
\left(
\begin{array}{cccc}
0     &\ve2   &\Phb31 &0      \\
0     &\ve5   &\Ph23  &0      \\
\Ph31 &\Phb23 &0/\ve7      &\phb45 \\
0     &\ve9   &\ph45  &0      \\
\end{array}
\right)
$
&
$\left[\begin{array}{c}
\cph45\cve5-\cPh23\cve9\\
\cPhb31\cve9-\cph45\cve2\\
0\\
\cPh23\cve2-\cPhb31\ve5\\
\end{array}
\right]
$
& 
$\left[\begin{array}{c}
\phantom{-}\phb45\\
0\\
0\\
-\Ph31
\end{array}
\right]
$
&
\begin{minipage}{4cm}
$(\phb45 \lb1-\Ph31\lb4)$\\
$\times (\ph45 \ve5\l1-\Ph23\ve9\l1$\\
$+\Ph23\ve2\l4-\Phb31\ve5\l4$\\
$-\ph45\ve2\l2+\Phb31\ve9\l2)$
\end{minipage}
\\\hline
\end{tabular}
\end{center}
\rightline{\hfil Table \ref{oht}\ {\it (continued)}}
\end{table}
\begin{table}
\begin{center}
\begin{tabular}{|l|c|c|c|c|}
\hline
20&
$
\left(
\begin{array}{cccc}
0     &\ve2   &\Phb31 &\ve3      \\
0     &0      &\Ph23  &0      \\
\Ph31 &\Phb23 &0/\ve7      &\phb45 \\
0     &0      &\ph45  &0      \\
\end{array}
\right)
$
&
$\left[\begin{array}{c}
\phantom{-}0\\
\phantom{-}\cph45\\
\phantom{-}0\\
-\cPh23\\
\end{array}
\right]
$
& 
$\left[\begin{array}{c}
\Phb23\ve3-\phb45\ve2\\
-\Ph31\ve3\\
0\\
\Ph31\ve2
\end{array}
\right]
$
&
\begin{minipage}{4cm}
$(\Ph23\l4-\ph45\l2)$\\
$\times (\phb45\ve2\lb1-\Phb23\ve3\lb1$\\
$+\Ph31\ve3\lb2-\Ph31\ve2\lb4)$
\end{minipage}
\\\hline
21&
$
\left(
\begin{array}{cccc}
\ve1  &0      &\Phb31 &0      \\
0     &0      &\Ph23  &0      \\
\Ph31 &\Phb23 &0/\ve7      &\phb45 \\
0     &0      &\ph45  &0      \\
\end{array}
\right)
$
&
$\left[\begin{array}{c}
\phantom{-}0\\
\phantom{-}\cph45\\
\phantom{-}0\\
-\cPh23\\
\end{array}
\right]
$
& 
$\left[\begin{array}{c}
0\\
\phantom{-}\phb45\\
0\\
-\Phb23\\
\end{array}
\right]
$
&
\begin{minipage}{4cm}
$\ve1(\ph45\l2-\Ph23\l4)$\\
$\times (\phb45\lb2-\Phb23\lb4)$
\end{minipage}
\\\hline
22&
$
\left(
\begin{array}{cccc}
\ve1  &       &\Phb31 &0      \\
0     &0      &\Ph23  &0      \\
\Ph31 &\Phb23 &0/\ve7      &\phb45 \\
\ve8  &0      &\ph45  &0      \\
\end{array}
\right)
$
&
$\left[\begin{array}{c}
\cPh23\cve8\\
\cph45\cve1-\cPhb31\cve8\\
0\\
-\cPh23\cve1\\
\end{array}
\right]
$
& 
$\left[\begin{array}{c}
0\\
\phantom{-}\phb45\\
0\\
-\Phb23\\
\end{array}
\right]
$
&
\begin{minipage}{4cm}
$(\phb45\lb2-\Phb23\lb4)$\\
$\times (\ph45\ve1\l2-\Phb31\ve8\l2$\\
$+\Ph23\ve8\l1-\Ph23\ve1\l4)$
\end{minipage}
\\\hline
23&
$
\left(
\begin{array}{cccc}
\ve1  &0      &\Phb31 &0      \\
\ve4  &0      &\Ph23  &0      \\
\Ph31 &\Phb23 &0/\ve7      &\phb45 \\
0     &0      &\ph45  &0      \\
\end{array}
\right)
$
&
$\left[\begin{array}{c}
\cph45\cve4\\
-\cph45\cve1\\
\phantom{-}0\\
\cPh23\cve1-\cPhb31\cve4\\
\end{array}
\right]
$
& 
$\left[\begin{array}{c}
0\\
\phantom{-}\phb45\\
0\\
-\Phb23\\
\end{array}
\right]
$
&
\begin{minipage}{4cm}
$(-\Phb23\lb4+\phb45\lb2)$\\
$\times (\ph45\ve4\l1-\ph45\ve1\l2$\\
$+\Ph23\ve1\l4-\Phb31\ve4\l4)$
\end{minipage}
\\\hline
24&
$
\left(
\begin{array}{cccc}
\ve1     &0      &\Phb31 &0      \\
\ve4     &0      &\Ph23  &0      \\
\Ph31 &\Phb23 &0/\ve7      &\phb45 \\
\ve8     &0      &\ph45  &0      \\
\end{array}
\right)
$
&
$\left[\begin{array}{c}
\cph45\cve4-\cPh23\cve8\\
\cPhb31\cve8-\cph45\cve1\\
\phantom{-}0\\
\cPh23\cve1-\cPhb31\cve4\\
\end{array}
\right]
$
& 
$\left[\begin{array}{c}
0\\
\phantom{-}\phb45\\
0\\
-\Phb23\\
\end{array}
\right]
$
&
\begin{minipage}{4cm}
$(\Phb23\lb4-\phb45\lb2)$\\
$\times (\ph45\ve4\l1-\Ph23\ve8\l1$\\
$-\ph45\ve1\l2+\Phb31\ve8\l2$\\
$+\Ph23\ve1\l4-\Phb31\ve4\l4)$
\end{minipage}
\\\hline
25&
$
\left(
\begin{array}{cccc}
\ve1     &0   &\Phb31 &\ve3      \\
0     &0      &\Ph23  &0      \\
\Ph31 &\Phb23 &0/\ve7      &\phb45 \\
0     &0      &\ph45  &0      \\
\end{array}
\right)
$
&
$\left[\begin{array}{c}
\phantom{-}0\\
\phantom{-}\cph45\\
\phantom{-}0\\
-\cPh23\\
\end{array}
\right]
$
& 
$\left[\begin{array}{c}
\Phb23\ve3\\
\phb45\ve1 -\Ph31\ve3\\
0\\
-\Phb23\ve1\\
\end{array}
\right]
$
&
\begin{minipage}{4cm}
$(\ph45\l2-\Ph23\l4)$\\
$\times (\phb45\ve1\lb2-\Ph31\ve3\lb2$\\
$+\Phb23\ve3\lb1-\Phb23\ve1\lb4)$
\end{minipage}
\\\hline
26&
$
\left(
\begin{array}{cccc}
\ve1     &\ve2   &\Phb31 &0      \\
0     &0      &\Ph23  &0      \\
\Ph31 &\Phb23 &0/\ve7      &\phb45 \\
0     &0      &\ph45  &0      \\
\end{array}
\right)
$
&
$\left[\begin{array}{c}
\phantom{-}0\\
\phantom{-}\cph45\\
\phantom{-}0\\
-\cPh23\\
\end{array}
\right]
$
& 
$\left[\begin{array}{c}
\phb45\ve2\\
-\phb45\ve1\\
0\\
\Phb23\ve1-\Ph31\ve2\\
\end{array}
\right]
$
&
\begin{minipage}{4cm}
$(\Ph23\l4-\ph45\l2)$\\
$\times (\phb45\ve2\lb1-\phb45\ve1\lb2$\\
$+\Phb23\ve1\lb4-\Ph31\ve2\lb4)$
\end{minipage}
\\\hline
27&
$
\left(
\begin{array}{cccc}
\ve1     &\ve2   &\Phb31 &\ve3      \\
0     &0      &\Ph23  &0      \\
\Ph31 &\Phb23 &0/\ve7      &\phb45 \\
0     &0      &\ph45  &0      \\
\end{array}
\right)
$
&
$\left[\begin{array}{c}
\phantom{-}0\\
\phantom{-}\cph45\\
\phantom{-}0\\
-\cPh23\\
\end{array}
\right]
$
& 
$\left[\begin{array}{c}
\phb45\ve2-\Phb23\ve3\\
\Ph31\ve3-\phb45\ve1\\
0\\
\Ph23\ve1-\Ph31\ve2
\end{array}
\right]
$
&
\begin{minipage}{4cm}
$(\Ph23\l4-\ph45\l2)$\\
$\times (\phb45\ve2\lb1-\Phb23\ve3\lb1$\\
$-\phb45\ve1\lb2+\Phb23\ve1\lb4$\\
$+\Ph31\ve3\lb2-\Ph31\ve2\lb4)$
\end{minipage}
\\\hline
\end{tabular}
\end{center}
\rightline{\hfil Table \ref{oht}\ {\it (continued)}}
\end{table}
\begin{table}
\begin{eqnarray}
&&   {{h}_{1}} {\overline{h}_{1}} {{D}_{5}} {{D}_{5}} {\overline{\phi}_{3}} {\overline{\phi}_{3}} {{\Phi}_{23}}+
   {{h}_{1}} {\overline{h}_{1}} {{D}_{5}} {{D}_{5}} {\overline{\phi}_{4}} {\overline{\phi}_{4}} {{\Phi}_{23}}+
   {{h}_{1}} {\overline{h}_{1}} {{D}_{5}} {{T}_{4}} {{D}_{2}} {{T}_{2}} {{\phi}_{2}}+ {{h}_{1}} {\overline{h}_{1}} {{T}_{5}} {{T}_{5}} {{\phi}_{2}} {{\phi}_{2}} {\overline{\Phi}_{23}}+
\nonumber\\&&
   {{h}_{1}} {\overline{h}_{1}} {{T}_{5}} {{T}_{5}} {{\phi}_{4}} {{\phi}_{4}} {\overline{\Phi}_{23}}+
   {{h}_{1}} {\overline{h}_{1}} {{D}_{2}} {{D}_{2}} {{\phi}_{2}} {{\phi}_{2}} {\overline{\Phi}_{23}}+
   {{h}_{1}} {\overline{h}_{1}} {{D}_{2}} {{D}_{2}} {{\phi}_{4}} {{\phi}_{4}} {\overline{\Phi}_{23}}+
   {{h}_{1}} {\overline{h}_{1}} {{T}_{2}} {{T}_{2}} {{\phi}_{2}} {{\phi}_{2}} {\overline{\Phi}_{23}}+
\nonumber\\&&
   {{h}_{1}} {\overline{h}_{1}} {{T}_{2}} {{T}_{2}} {{\phi}_{4}} {{\phi}_{4}} {\overline{\Phi}_{23}}+
   {{h}_{1}} {\overline{h}_{2}} {{D}_{5}} {{D}_{4}} {\overline{\phi}_{3}} {\overline{\phi}_{3}} {\overline{\phi}_{3}}+
   {{h}_{1}} {\overline{h}_{2}} {{D}_{5}} {{D}_{4}} {\overline{\phi}_{3}} {\overline{\phi}_{4}} {\overline{\phi}_{4}}+
   {{h}_{1}} {\overline{h}_{2}} {{T}_{5}} {{T}_{4}} {\overline{\phi}_{3}} {\overline{\phi}_{3}} {{\phi}_{2}}+
\nonumber\\&&
   {{h}_{1}} {\overline{h}_{2}} {{T}_{5}} {{T}_{4}} {{\phi}_{2}} {\overline{\phi}_{4}} {\overline{\phi}_{4}}+
   {{h}_{1}} {\overline{h}_{45}} {{T}_{5}} {{T}_{5}} {\overline{\phi}_{45}}+
   {{h}_{1}} {\overline{h}_{45}} {{D}_{2}} {{D}_{2}} {\overline{\phi}_{45}}+
   {{h}_{1}} {\overline{h}_{45}} {{T}_{2}} {{T}_{2}} {\overline{\phi}_{45}}+
   {{h}_{1}} {\overline{h}_{45}} {{D}_{5}} {{T}_{5}} {{D}_{3}} {{T}_{3}} {\overline{\phi}_{3}}+
\nonumber\\&&
   {{h}_{1}} {\overline{h}_{45}} {{D}_{5}} {{T}_{5}} {{D}_{2}} {{T}_{2}} {\overline{\phi}_{45}}+
   {{h}_{1}} {\overline{h}_{45}} {{D}_{5}} {{D}_{4}} {\overline{\phi}_{3}} {\overline{\phi}_{45}} {\overline{\Phi}_{31}}+
   {{h}_{1}} {\overline{h}_{45}} {{T}_{5}} {{T}_{5}} {\overline{\phi}_{45}} {{\Phi}_{23}} {\overline{\Phi}_{23}}+
   {{h}_{1}} {\overline{h}_{45}} {{T}_{5}} {{D}_{4}} {{D}_{3}} {{T}_{3}} {\overline{\Phi}_{23}}+
\nonumber\\&&
   {{h}_{1}} {\overline{h}_{45}} {{T}_{5}} {{T}_{4}} {{\phi}_{2}} {\overline{\phi}_{45}} {\overline{\Phi}_{31}}+
   {{h}_{1}} {\overline{h}_{45}} {{D}_{4}} {{D}_{4}} {\overline{\phi}_{3}} {\overline{\phi}_{3}} {\overline{\phi}_{45}}+
   {{h}_{1}} {\overline{h}_{45}} {{D}_{4}} {{D}_{4}} {\overline{\phi}_{4}} {\overline{\phi}_{4}} {\overline{\phi}_{45}}+
   {{h}_{1}} {\overline{h}_{45}} {{D}_{2}} {{D}_{2}} {\overline{\phi}_{45}} {{\Phi}_{23}} {\overline{\Phi}_{23}}+
\nonumber\\&&
   {{h}_{1}} {\overline{h}_{45}} {{T}_{2}} {{T}_{2}} {\overline{\phi}_{45}} {{\Phi}_{23}} {\overline{\Phi}_{23}}+
   {{h}_{2}} {\overline{h}_{1}} {{D}_{5}} {{D}_{4}} {\overline{\phi}_{3}} {{\phi}_{2}} {{\phi}_{2}}+
   {{h}_{2}} {\overline{h}_{1}} {{T}_{5}} {{T}_{4}} {{\phi}_{2}} {{\phi}_{2}} {{\phi}_{2}}+
\nonumber\\&&
   {{h}_{2}} {\overline{h}_{1}} {{T}_{5}} {{T}_{4}} {{\phi}_{2}} {{\phi}_{4}} {{\phi}_{4}}+
   {{h}_{2}} {\overline{h}_{2}} {{D}_{4}} {{D}_{4}} {\overline{\phi}_{3}} {\overline{\phi}_{3}} {{\Phi}_{31}}+
   {{h}_{2}} {\overline{h}_{2}} {{D}_{4}} {{D}_{4}} {\overline{\phi}_{4}} {\overline{\phi}_{4}} {{\Phi}_{31}}+
   {{h}_{2}} {\overline{h}_{2}} {{T}_{4}} {{T}_{4}} {{\phi}_{2}} {{\phi}_{2}} {\overline{\Phi}_{31}}+
\nonumber\\&&
   {{h}_{2}} {\overline{h}_{2}} {{T}_{4}} {{T}_{4}} {{\phi}_{4}} {{\phi}_{4}} {\overline{\Phi}_{31}}+
   {{h}_{2}} {\overline{h}_{45}} {{D}_{4}} {{D}_{4}} {\overline{\phi}_{45}}+
   {{h}_{2}} {\overline{h}_{45}} {{D}_{5}} {{D}_{4}} {\overline{\phi}_{3}} {\overline{\phi}_{45}} {{\Phi}_{23}}+
   {{h}_{2}} {\overline{h}_{45}} {{T}_{5}} {{T}_{5}} {{\phi}_{2}} {{\phi}_{2}} {\overline{\phi}_{45}}+
\nonumber\\&&
   {{h}_{2}} {\overline{h}_{45}} {{T}_{5}} {{T}_{5}} {{\phi}_{4}} {{\phi}_{4}} {\overline{\phi}_{45}}+
   {{h}_{2}} {\overline{h}_{45}} {{T}_{5}} {{D}_{4}} {{D}_{3}} {{T}_{3}} {{\Phi}_{31}}+
   {{h}_{2}} {\overline{h}_{45}} {{T}_{5}} {{T}_{4}} {{\phi}_{2}} {\overline{\phi}_{45}} {{\Phi}_{23}}+
   {{h}_{2}} {\overline{h}_{45}} {{D}_{4}} {{D}_{4}} {\overline{\phi}_{45}} {\overline{\Phi}_{31}} {{\Phi}_{31}}+
\nonumber\\&&
   {{h}_{2}} {\overline{h}_{45}} {{D}_{4}} {{T}_{4}} {{D}_{3}} {{T}_{3}} {{\phi}_{2}}+
   {{h}_{2}} {\overline{h}_{45}} {{D}_{2}} {{D}_{2}} {{\phi}_{2}} {{\phi}_{2}} {\overline{\phi}_{45}}+
   {{h}_{2}} {\overline{h}_{45}} {{D}_{2}} {{D}_{2}} {{\phi}_{4}} {{\phi}_{4}} {\overline{\phi}_{45}}+
   {{h}_{2}} {\overline{h}_{45}} {{T}_{2}} {{T}_{2}} {{\phi}_{2}} {{\phi}_{2}} {\overline{\phi}_{45}}+
\nonumber\\&&
   {{h}_{2}} {\overline{h}_{45}} {{T}_{2}} {{T}_{2}} {{\phi}_{4}} {{\phi}_{4}} {\overline{\phi}_{45}}+
   {{h}_{45}} {\overline{h}_{1}} {{D}_{5}} {{D}_{5}} {{\phi}_{45}}+
   {{h}_{45}} {\overline{h}_{1}} {{D}_{5}} {{D}_{5}} {{\phi}_{45}} {{\Phi}_{23}} {\overline{\Phi}_{23}}+
   {{h}_{45}} {\overline{h}_{1}} {{D}_{5}} {{D}_{4}} {\overline{\phi}_{3}} {{\phi}_{45}} {{\Phi}_{31}}+
\nonumber\\&&
   {{h}_{45}} {\overline{h}_{1}} {{T}_{5}} {{T}_{4}} {{\phi}_{2}} {{\phi}_{45}} {{\Phi}_{31}}+
   {{h}_{45}} {\overline{h}_{1}} {{T}_{4}} {{T}_{4}} {{\phi}_{2}} {{\phi}_{2}} {{\phi}_{45}}+
   {{h}_{45}} {\overline{h}_{1}} {{T}_{4}} {{T}_{4}} {{\phi}_{4}} {{\phi}_{4}} {{\phi}_{45}}+
   {{h}_{45}} {\overline{h}_{2}} {{T}_{4}} {{T}_{4}} {{\phi}_{45}}+
\nonumber\\&&
   {{h}_{45}} {\overline{h}_{2}} {{D}_{5}} {{D}_{5}} {\overline{\phi}_{3}} {\overline{\phi}_{3}} {{\phi}_{45}}+
   {{h}_{45}} {\overline{h}_{2}} {{D}_{5}} {{D}_{5}} {\overline{\phi}_{4}} {\overline{\phi}_{4}} {{\phi}_{45}}+
   {{h}_{45}} {\overline{h}_{2}} {{D}_{5}} {{D}_{4}} {\overline{\phi}_{3}} {{\phi}_{45}} {\overline{\Phi}_{23}}+
   {{h}_{45}} {\overline{h}_{2}} {{T}_{5}} {{T}_{4}} {{\phi}_{2}} {{\phi}_{45}} {\overline{\Phi}_{23}}+
\nonumber\\&&
   {{h}_{45}} {\overline{h}_{2}} {{T}_{4}} {{T}_{4}} {{\phi}_{45}} {\overline{\Phi}_{31}} {{\Phi}_{31}}+
   {{h}_{45}} {\overline{h}_{45}} {{D}_{5}} {{D}_{5}} {{\phi}_{4}} {\overline{\phi}_{4}} {\overline{\Phi}_{31}}+
   {{h}_{45}} {\overline{h}_{45}} {{D}_{5}} {{D}_{5}} {{\phi}_{45}} {\overline{\phi}_{45}} {\overline{\Phi}_{31}}+
\nonumber\\&&
   {{h}_{45}} {\overline{h}_{45}} {{D}_{5}} {{D}_{4}} {\overline{\phi}_{3}} {{\phi}_{45}} {\overline{\phi}_{45}}+
   {{h}_{45}} {\overline{h}_{45}} {{T}_{5}} {{T}_{5}} {{\phi}_{4}} {\overline{\phi}_{4}} {{\Phi}_{31}}+
   {{h}_{45}} {\overline{h}_{45}} {{T}_{5}} {{T}_{5}} {{\phi}_{45}} {\overline{\phi}_{45}} {{\Phi}_{31}}+
   {{h}_{45}} {\overline{h}_{45}} {{T}_{5}} {{T}_{4}} {{\phi}_{2}} {{\phi}_{4}} {\overline{\phi}_{4}}+
\nonumber\\&&
   {{h}_{45}} {\overline{h}_{45}} {{T}_{5}} {{T}_{4}} {{\phi}_{2}} {{\phi}_{45}} {\overline{\phi}_{45}}+
   {{h}_{45}} {\overline{h}_{45}} {{D}_{4}} {{D}_{4}} {{\phi}_{4}} {\overline{\phi}_{4}} {\overline{\Phi}_{23}}+
   {{h}_{45}} {\overline{h}_{45}} {{D}_{4}} {{D}_{4}} {{\phi}_{45}} {\overline{\phi}_{45}} {\overline{\Phi}_{23}}+
   {{h}_{45}} {\overline{h}_{45}} {{T}_{4}} {{T}_{4}} {{\phi}_{4}} {\overline{\phi}_{4}} {{\Phi}_{23}}+
\nonumber\\&&
   {{h}_{45}} {\overline{h}_{45}} {{T}_{4}} {{T}_{4}} {{\phi}_{45}} {\overline{\phi}_{45}} {{\Phi}_{23}}+
   {{h}_{45}} {\overline{h}_{45}} {{D}_{2}} {{D}_{2}} {{\phi}_{4}} {\overline{\phi}_{4}} {{\Phi}_{31}}+
   {{h}_{45}} {\overline{h}_{45}} {{D}_{2}} {{D}_{2}} {{\phi}_{45}} {\overline{\phi}_{45}} {{\Phi}_{31}}+
   {{h}_{45}} {\overline{h}_{45}} {{T}_{2}} {{T}_{2}} {{\phi}_{4}} {\overline{\phi}_{4}} {{\Phi}_{31}}+
\nonumber\\&&
   {{h}_{45}} {\overline{h}_{45}} {{T}_{2}} {{T}_{2}} {{\phi}_{45}} {\overline{\phi}_{45}} {{\Phi}_{31}}
\end{eqnarray}
\caption{{\it Non--renormalizable contributions to the Higgs doublet 
mass matrix that are non--zero up
to seventh order for the vacuum choice (\ref{zeroes}).}}
\end{table}


\newpage

\begin{thebibliography}{99}

\bibitem{DG} S.Dimopoulos and H. Georgi, Nucl. Phys. {\bf B193} (1981)
150.

\bibitem{TDA} S. Dimopoulos and F. Wilczek, in the {\it Proc. of the 
International School on Subnuclear Physics}, Erice, Italy,
1981.

\bibitem{TDB} A. Masiero, D.V. Nanopoulos, K. Tamvakis and T.
Yanagida, Phys. Lett. {\bf B115} (1982) 380.

\bibitem{FSU5B}
I. Antoniadis, J. Ellis, J. Hagelin and D.V. Nanopoulos,
Phys.Lett. {\bf B194} (1987) 231.

\bibitem{FSU5A} S. M. Barr, Phys. Lett. {\bf B112} (1982) 219;\\ 
J. P. Deredinger, J. E. Kim and D. V. Nanopoulos, Phys. Lett. {\bf B139} (1984) 
170.

\bibitem{aehn} I. Antoniadis, J. Ellis, J. Hagelin and D.V. Nanopoulos,
Phys. Lett. {\bf B231} (1989) 65.

\bibitem{NR}S. Kalara, J. L. Lopez and D.V. Nanopoulos,
Phys. Lett. {\bf B245} (1990) 421.

\bibitem{ELLN} J. Ellis, G. K. Leontaris, S. Lola and D.V. Nanopoulos,
Phys. Lett. {\bf B425} (1998) 86 and Eur. Phys. J. {\bf C9} (1999) 389.

\bibitem{KTJR} J. Rizos and K. Tamvakis, Phys. Lett. {\bf B251} (1990) 369.

\bibitem{ART} I. Antoniadis, J. Rizos and K. Tamvakis, Phys. Lett.
{\bf B278} (1992) 257.

\end{thebibliography}
\end{document}